# Web Based Reputation Index of Turkish Universities

Mehmet Lutfi ARSLAN, Sadi Evren SEKER

*Abstract*—**This paper attempts to develop an online reputation index of Turkish universities through their online impact and effectiveness. Using 16 different web based parameters and employing normalization process of the results, we have ranked websites of Turkish universities in terms of their web presence. This index is first attempt to determine the tools of reputation of Turkish academic websites and would be a basis for further studies to examine the relation between reputation and the online effectiveness of the universities.**

*Index Terms*—**Online Reputation, Webometrics, Web Mining, Higher Education.**

Websites have become main venues for reputation and presence of universities. Indeed, a web site of a university is vital medium for promoting the universities core competencies. It is an indicator of how this institution is perceived by its internal and external customers, reflecting not only its academic performance, but also its administrative services. Thus, the Internet and dissemination of information is forcing universities to have more visible and effective web presence.

Along with the global interest in university rankings, impact and effectiveness of university web sites have become subject to numerous studies and projects. One of them is the project utilized by the Cybermetrics Lab, a research group of the Spanish National Research Council (CSIC). This Project is called "The Webometrics Ranking of World Universities" based on university web presence, visibility and web access. As it is stated in the project's web site, "the objective is not to evaluate websites, their design or usability or the popularity of their contents according to the number of visits or visitors." The ranking's first edition was published in 2004. Since 2006, it appears twice per year [1].

Using web mining techniques such as data extracted from search and counting number of inlinks, it is possible to generate new ranking methods. This paper is an attempt to do so by employing a bunch of indicators to measure the impact of academic web sites.

In this paper, we have tried to create a reputation index by the web indicators like Google page rank, number of visitors, number of pages linking back to the web page or the number of likes on Facebook.

We have collected 16 different web indicator data and normalized the collected numbers using min-max normalization. After normalizing all the collected data, we have got an index value for all of the universities via arithmetic mean. Finally the university index values are normalized via the number of students attending to the university, because the web index of universities are closely related to the number of students which also correlated with the number of staff in the university. The normalized university web index is calculated by dividing the university web index to the total number of students attending to the university.

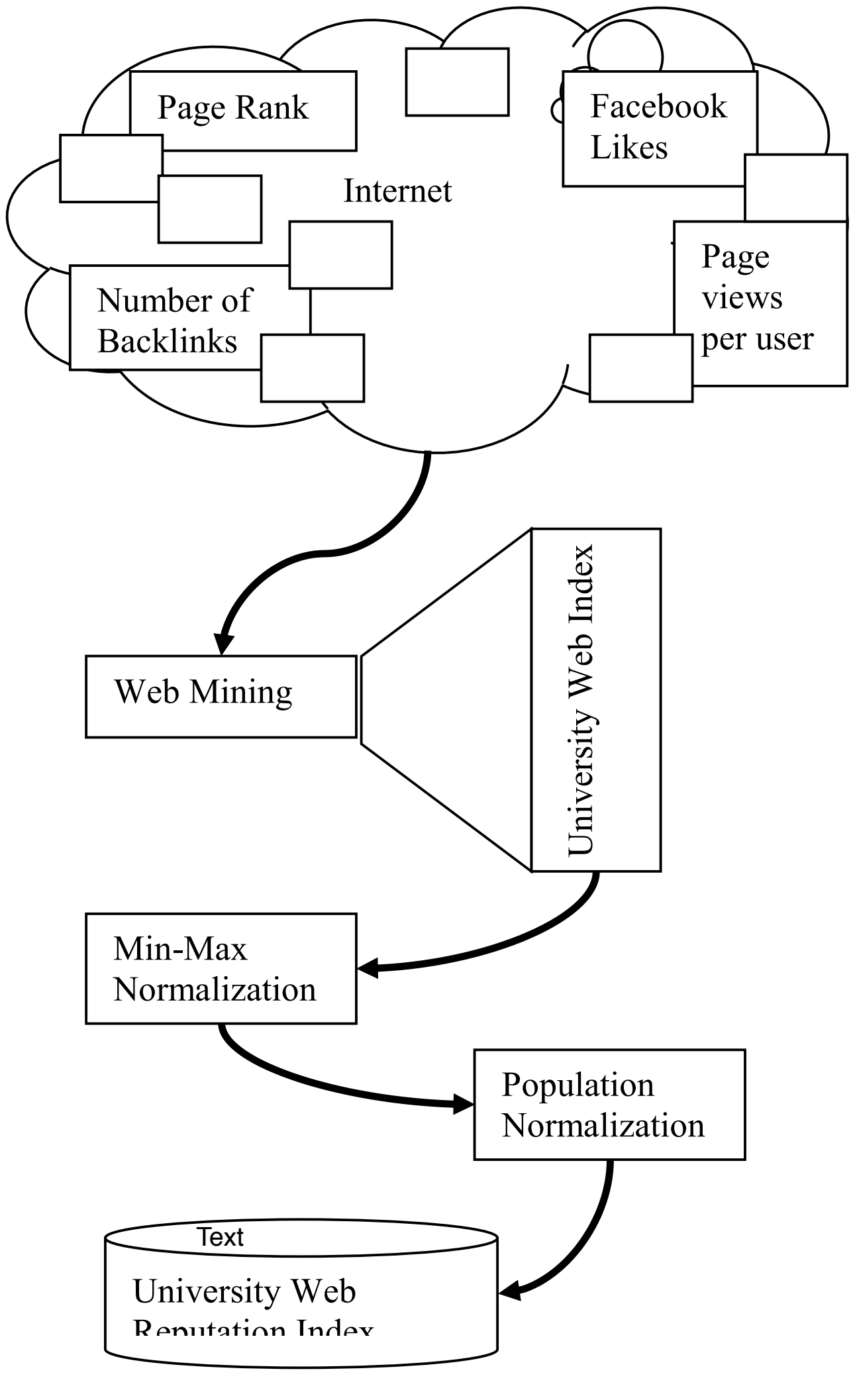

Fig. 1. Data Flow Dıagram of the Study.

As it is demonstrated on Figure 1, the web indicators of a web page of all the universities are gathered from the internet resources. The first gathered information is then passed through two more steps which are min-max normalization and the population normalization. Finally a university web reputation index is achieved and the universities are ranked via this index first time on this study.

M. L. Arslan is with Business Administration Department, Istanbul Medeniyet University, Istanbul, Turkey (e-mail: lutfi.arslan@medeniyet. edu.tr).

S. E. Şeker was with Computer Department, Istanbul University, Istanbul, Turkey. He is now Business Administration Department, Istanbul Medeniyet University, Istanbul, Turkey (sadi.seker@medeniyet. edu.tr).

## I. Background

Employing specific web mining techniques, researchers try to measure the impact and effectiveness of academic web sites. Quantitative aspects of web data and information have created separate sub-disciplines like informetrics, cybermetrics, and webometrics. Informetrics is "the study of the quantitative aspects of information in any form, not just records or bibliographies, and in any social group, not just scientists" [2], and cybermetrics is "the study of the quantitative aspects of the construction and use of information resources, structures and technologies on the whole Internet drawing on bibliometric and informetric approaches" [3].

Webometrics, a term first coined by Almind and Ingwersen, is a measurement of the effectiveness of web sites [4]. According to Thelwall, it is "the study of web-based content with primarily quantitative methods for social science research goals using techniques that are not specific to one field of study" [5]. First example of this measurement is the "Web Impact Factor" (WIF) developed by Ingwersen and defined as "the number of web pages in a web site receiving links from other web sites, divided by the number of web pages published in the site that are accessible to the crawler" [6].

Similar techniques, which have been developed throughout the years, all showed that there is a significant correlation between academic performance and impact of web sites of the universities [7], [8], [9], [10]. Especially, a study by Qiu, Chen, and Wang showed that external inlinks correlated general ratings of the universities [11]. Another study by Aguillo, Granadino, Ortega, and Prieto on 9,330 institutions worldwide, found significant correlations between webometric data and bibliometric data [12]. For business companies, same kind of relation was also found. There is a significant correlation between the number of inlinks to the web site of a company and its business performance [13], [14].

The proposition behind webometrics is that web visibility and impact of a university is highly correlated with its reputation. Those universities with good reputation tend to have more visible web sites, high traffic, more links etc.

There are five set of tools of webometric research: link analysis, web citation analysis, search engine evaluation, descriptive studies of the web, and the analysis of Web 2.0 phenomena. While link analysis measures the hyperlinks between web pages, web citation analysis counts how often journal articles are cited. Search engines are used to evaluate the extent of the coverage of the web and the accuracy of the reported results. Descriptive studies include various survey methods like the average web page size, average number and type of meta-tags used, the average use of technologies like Java and JavaScript, the number of users, pages and web servers. Last but not least tool is Web 2.0 applications [15].

As the aim of our paper, we use tools of webometric research like Google page rank, number of visitors, number of pages linking back to the web page or the number of likes on Facebook, in order to create a reputation index. Our intention is to be as simple and usable as possible. We can summarize our method as follows:

## II. Mining

In the web mining phase, we have collected 16 different indicators of a web page from the web resources. The indicators and brief explanations are placed in this section.

Before getting into the details of the indicators, we should indicate that the number of universities in Turkey is 170 and some of them are relatively new like about only a few years. Obviously the index is created on the current status of the universities and a change can be expected by the time.

**Has a Facebook Page?** We consider the social media as a part of the web reputation and we have checked if the university has a Facebook fan page or not. Only 128 of the universities among 170 have a Facebook page.

**Facebook Like Count**. Another indicator is the number of likes of the Facebook fan page. Among the 128 universities who has a Facebook page, the maximum number of likes is 71114 and the average like count is 7817.

**Value of the Site**. Some of the independent organizations offers a free agent to calculate the expected value of the web site via the web indicators like Alexa ranking or Google page rank. Most of them are built on the number of visitors and expected click from the visitors to make a valuing. The maximum expected value of a university web page in Turkey is 326.642 and average value of the university sites in Turkey is 11.965 USD.

**Yahoo! Backlinks**. The number of backlinks is provided by Yahoo!. Depending on the crawler of the Yahoo!, the backlinks are counted by the number of other web pages holding links to the university web page. The maximum number of Yahoo! backlinks is 80400 and the average number of back links is 7749.

**Google Backlinks**. The number of different web pages holding a link to the university web page depending on the Google web crawler. We need to indicate that the number of backlinks counted by Google is much more higher than the yahoo crawler. This might happen because of deeper crawling by Google. The maximum number of back links is 84 million while the average is 17million per university web page.

**DMOZ Index** is a open directory project where the web pages are listed via their categories. We have checked if the web pages of the universities are listed in DMOZ (Directory MOZilla). This indicator has no effect on the result because all of the universities have a DMOZ listing.

**Number of Google Indexed Pages.** Another indicator is the number of web pages indexed by Google. The web pages of the universities have different number of subpages. For example the maximum number of Google indexed page is about 9 million while the average number of Google indexed page is 234.883. The average number of Google indexed page indicates that an average university may hold about 234.883 web pages in the same domain which may be the home pages of the professors, the student pages, the course pages, the administrative pages or the announcements.

**Yahoo Indexed Pages** is another indicator in our indexing study. The Yahoo indexed page count is relatively less than the Google count, where the maximum is about 4.400.000 and the average is 34.327.

**Daily Unique Visitors** is the average number of visitors per day. Also this number is seasonal and can change from season to season. In order to make a clean room measurement, we have collected the information during the off season, which is the summer term and most of the universities has no attendants. As expected the number of unique visitors is increasing during the final and registration weeks and since all the universities have different calendar we have collected the information during the off-season. The maximum visitor is 19.718 and the average is 757.

**Plagiarism** is another web indicator, where there are online web robots trying to find similarities on the web site content and other resources. These similarities cannot be considered as academic plagiarism in most of the time. For example a paper published by a professor is both published in his personal home page on the university web page and another academic directory at the same time. These robots can consider this type of duplicate publishing as a plagiarism. Although the plagiarism robots are not good in finding an academic plagiarism in most of the time, they are quite good in indicating the number of fresh publishing from the university academic staff. We have included the number of plagiarisms returned for each of the university web page as another indicator in our study and the maximum value is 10 while the average is 4.

**Speed Test**. We have also executed some speed tests from 5 different global locations. The speed tests are simply executed by the ping rates and the average of 5 ping rate is normalized with a higher weight to the speed test from Turkey since most of the visitors are connecting those sites from Turkey. The ping rates are also gathered in different days times in order to avoid a temporary site failure. The maximum ping rate in average calculated is 3208 miliseconds and average is 240 ms.

**Alexa Ranking** is another indicator published by an Amazon owned web site alexa.com. The lesser number means the web page has a higher ranking and the minimum ranking for the Turkish university web pages is 924 and highest ranking is 26992405 among the whole web sites on the Internet. The rankings above 20 million can be considered as a fresh web site.

**Alexa Bounce Rating** is another rating collected from alexa.com and indicates the number of bounces which means the users just visit a single page and then leaves the web site. The higher rates lower the web reputation index while the lower values indicate a higher reputation value. The maximum value is 90% and minimum is about 7%.

**Page Views Per User** is another indicator to calculate the number of pages visited by a single user. The higher number means the user is spending more time to visit more pages and we consider this time spending as an indicator to a more attractive web site. The maximum is 6.40 and the average is 1.83.

**Time on Site** is a web indicator to measure the time spending of the users with a time interval of their entrance and exit. The higher time means a higher reputation for the web site and the maximum value of time spent on the web page is about 13 minutes and average is about 4 minutes. These time intervals are also daily, which means the time on site indicator is an average day based time spending on web page for each of the user.

**The number of Sites Linking In** is another indicator that the number of inner links is counted. The number of inner links can be reviewed from two different perspectives. The former view can be the bigger web sites, which are relatively holding higher reputation has a high number of inner links. The latter view can be the strongly connected web sites where the users can navigate from one page to other easily and therefore the reputation is higher again, has higher inner link count. The average is 908 and maximum is 9200 for this indicator.

## III. Normalization

In the normalization phase, the collected web indicator values are normalized via min-max normalization.

$$N_{Min\,Max}(x) = \frac{x - Min}{Max - Min}$$

The normalized value is calculated by the subtraction of the minimum value of the series from the sample and dividing the subtraction to the distance between minimum and maximum values of the series.

The reason of normalization is getting comparable values for each of the indicators. For example the number of Google backlinks is varying from 80 million to 2 million while the speed test is varying from 3.2 seconds to 240 mili seconds. In order to get a final value from all these indicators we need a common range. The min-max normalization value always results between 0 and 1.

Most of the indicators we have collected are have a positive impact on the web reputation while the normalized values are getting higher. The only exception of this fact is bounce rates and the Alexa rankings where the reputation gets worse while these numbers are increasing. As a solution we have calculated the inverse of these indicators by multiplying with -1. Which means a subtraction in the final decision in fact.

So the total score is calculated with below formula.

$$WRI = \frac{\sum_0^C N_x - \sum_C^K N_x}{C}$$

The Web Reputation Index (WRI) is calculated with the summation of negative indicators subtracted from the summation of positive indicators divided by the count of positive indicators “C”. The “K” symbol in above formula stands for the total number of indicators which is the summation of positive and negative indicator counts.

Because the summation of positive indicators is always higher than the summation of negative indicators the equation of WRI always gets a positive real number between 0 and 1.

Finally the WRI values of each university is divided into the min-max normalized population count of each university. The maximum is 73640 and minimum is 0 because there are some universities just founded this year. Obviously getting the 0 as a university population will yield a 0 normalized value after min-max normalization and dividing the WRI value to 0 will be a mathematically undefined problem. As a solution, we have considered the number of students as 1 for the new universities and calculated the population normalized values by dividing WRI to the population counts under this special predefinition.

$$N_{pop} = \frac{WRI}{P}, \begin{cases} P = 1\,, if\ population = 0 \\ P = popluation\,, if\ population > 0 \end{cases}$$

After the calculation of population normalized values we have a range of normalization between 0 and 1 again.

The results of index values after population normalized is given in the results section.

## IV. RESULTS

This section holds the details of the normalized index values. The complete list of universities with the index values are placed into the appendix of the paper.

Properties of the data set is given in Table I.

TABLE I: PROPERTIES OF THE INDEX VALUES

| | |
|---|---|
| Mean ( μ ) | 0.280 |
| Maximum | 0.449 |
| Minimum | 0.150 |
| Standard Deviation ( σ ) | 0.055 |
| Total Number of Universities | 170 |

The distribution of the university reputation index is given as a separate figure.

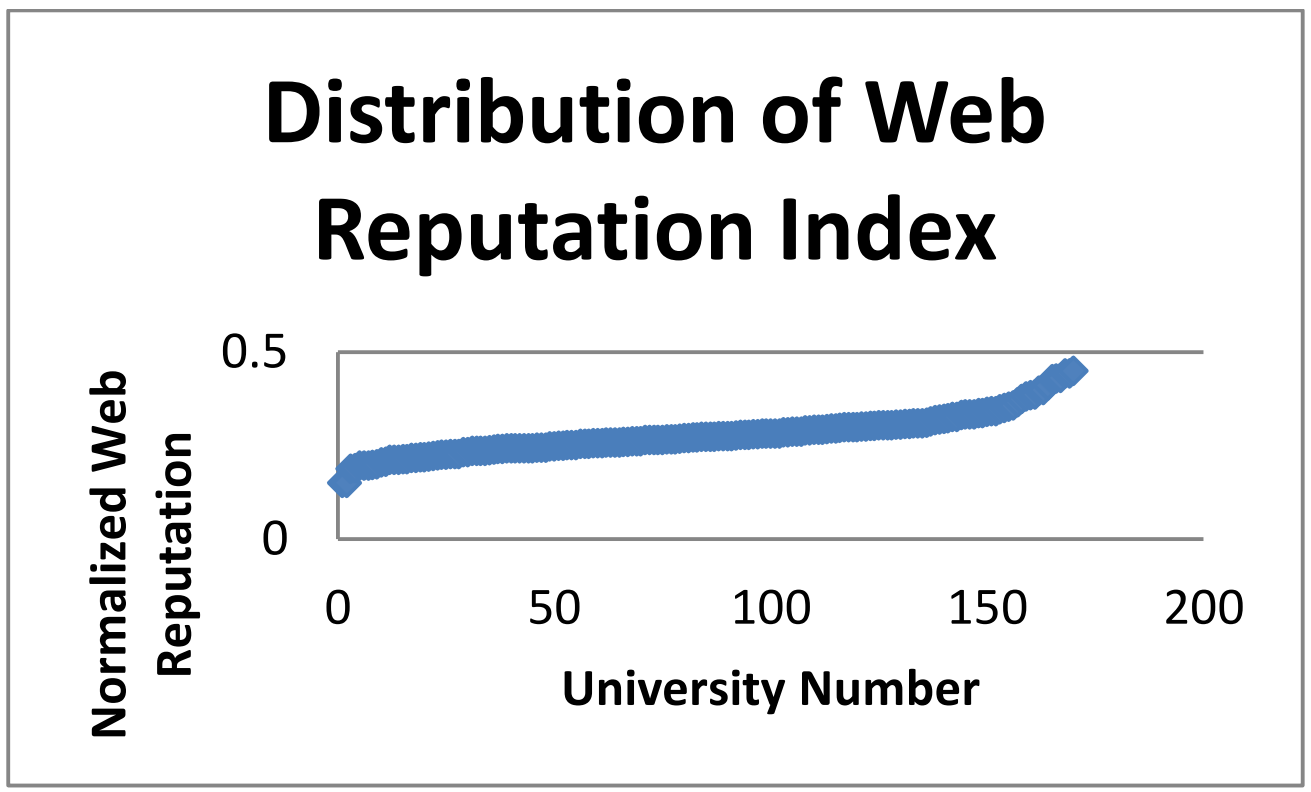

Fig. 2. Statistical distribution of normalized web reputation index

In figure 2, the x-axis holds a unique number for each of the university, while the y-axis demonstrates the normalized web reputation value.

## V. CONCLUSION

Employing 16 different web indicator data and normalizing the collected numbers using min-max normalization, we have ranked Turkish universities in terms of their online presence and impact. According to our results, ten most reputable and least reputable Turkish universities on the web are given in Table II, and Table III respectively.

TABLE II: THE MOST REPUTABLE TURKISH UNIVERSITIES ON THE WEB

| University | Value |
|---|---|
| Anadolu | 0,449508 |
| İstanbul | 0,444084 |
| Gazi | 0,443798 |
| Bilkent | 0,431946 |
| Sakarya | 0,42942 |
| Boğaziçi | 0,428301 |
| Hacettepe | 0,411034 |
| Ankara | 0,398226 |
| Yıldız Teknik | 0,397495 |
| Dokuz Eylül | 0,386229 |

TABLE III: THE LEAST REPUTABLE TURKISH UNIVERSITIES ON THE WEB

| University | Value |
|---|---|
| Deniz Harp Okulu | 0,150473 |
| Kara Harp Okulu | 0,151046 |
| Karatay | 0,188876 |
| Ankara Bilge | 0,19034 |
| Karabük | 0,196542 |
| Tunceli | 0,197094 |
| Şırnak | 0,197622 |
| K. Mehmetbey | 0,198129 |
| Avrasya | 0,200475 |
| Ağrı İ. Çeçen | 0,205266 |

This index of Turkish university websites is first attempt what proper tools of reputation could be. A further study could be a comparison between different ranking methods and to measure the validity of rankings respectively.

## APPENDIX

NORMALIZED WEB REPUTATION INDEX

| University | Normalized Index Value |
|---|---|
| Deniz Harp Okulu | 0,150473 |
| Kara Harp Okulu | 0,151046 |
| Karatay Üniversitesi | 0,188876 |
| Ankara Bilge Üniversitesi | 0,19034 |
| Karabük Üniversitesi | 0,196542 |
| Tunceli Üniversitesi | 0,197094 |
| Şırnak Üniversitesi | 0,197622 |
| Karamanoğlu Mehmetbey Üniversitesi | 0,198129 |
| Avrasya Üniversitesi | 0,200475 |
| Ağrı İbrahim Çeçen Üniversitesi | 0,205266 |
| Bozok Üniversitesi | 0,207951 |
| Kastamonu Üniversitesi | 0,211653 |
| Fatih Üniversitesi | 0,212482 |
| Nevşehir Üniversitesi | 0,212873 |
| Recep Tayyip Erdoğan Üniversitesi | 0,213605 |
| Iğdır Üniversitesi | 0,214206 |
| Amasya Üniversitesi | 0,216637 |
| Mimar Sinan Güzel Sanatlar Üniversitesi | 0,218165 |
| Uluslararası Antalya Üniversitesi | 0,219864 |
| Gülhane Askeri Tıp Akademisi | 0,219951 |
| Bingöl Üniversitesi | 0,220551 |
| Giresun Üniversitesi | 0,222696 |
| Canik Başarı Üniversitesi | 0,224975 |
| Üsküdar Üniversitesi | 0,225124 |
| Gümüşhane Üniversitesi | 0,225604 |
| İstanbul Bilim Üniversitesi | 0,226898 |
| Kafkas Üniversitesi | 0,227151 |
| İstanbul Gelişim Üniversitesi | 0,228233 |
| Dicle Üniversitesi | 0,232592 |
| Abdullah Gül Üniversitesi | 0,233152 |
| İpek Üniversitesi | 0,236428 |
| Türk Alman Üniversitesi | 0,236529 |
| İstanbul Medeniyet Üniversitesi | 0,236745 |
| Yalova Üniversitesi | 0,236952 |
| Gedik Üniversitesi | 0,23762 |
| Erzurum Teknik Üniversitesi | 0,239282 |
| Acıbadem Üniversitesi | 0,241863 |
| Muş Alparslan Üniversitesi | 0,241893 |
| Muğla Üniversitesi | 0,242247 |
| Karadeniz Teknik Üniversitesi | 0,242398 |
| Niğde Üniversitesi | 0,242484 |
| Cumhuriyet Üniversitesi | 0,24296 |
| Bayburt Üniversitesi | 0,24318 |
| Ufuk Üniversitesi | 0,243264 |
| İstanbul Medipol Üniversitesi | 0,243413 |
| Türk Hava Kurumu Üniversitesi | 0,244515 |
| İstanbul 29Mayıs Üniversitesi | 0,245219 |
| Yeni Yüzyıl Üniversitesi | 0,245316 |
| Erzincan Üniversitesi | 0,248421 |
| Siirt Üniversitesi | 0,248543 |
| Toros Üniversitesi | 0,249598 |
| İzmir Üniversitesi | 0,249888 |
| Trakya Üniversitesi | 0,251459 |
| İstanbul Arel Üniversitesi | 0,252223 |
| İstanbul Mef Üniversitesi | 0,252224 |
| Bitlis Eren Üniversitesi | 0,254953 |
| Namık Kemal Üniversitesi | 0,255088 |
| Fırat Üniversitesi | 0,255303 |
| Turgut Özal Üniversitesi | 0,2565 |
| Mevlana Üniversitesi | 0,257532 |
| Ahi Evran Üniversitesi | 0,257574 |
| Necmettin Erbakan Üniversitesi | 0,257948 |
| Sinop Üniversitesi | 0,257972 |
| İzmir Ekonomi Üniversitesi | 0,258956 |
| Kilis 7 Aralık Üniversitesi | 0,259193 |
| Bursa Orhangazi Üniversitesi | 0,259598 |
| Uşak Üniversitesi | 0,260207 |
| Bursa Teknik Üniversitesi | 0,261822 |
| Batman Üniversitesi | 0,26214 |
| Nuh Naci Yazgan Üniversitesi | 0,262467 |
| Süleyman Şah Üniversitesi | 0,264976 |
| Atılım Üniversitesi | 0,26501 |
| Gediz Üniversitesi | 0,265313 |
| Çağ Üniversitesi | 0,265444 |
| Gebze Yüksek Teknoloji Enstitüsü | 0,266152 |
| Ardahan Üniversitesi | 0,266803 |
| İstanbul Kemerburgaz Üniversitesi | 0,267016 |
| Sabancı Üniversitesi | 0,267462 |
| Erciyes Üniversitesi | 0,268713 |
| Adıyaman Üniversitesi | 0,269635 |
| Bartın Üniversitesi | 0,270835 |
| Dumlupınar Üniversitesi | 0,272041 |
| Kırklareli Üniversitesi | 0,272147 |
| Maltepe Üniversitesi | 0,273141 |
| Mardin Artuklu Üniversitesi | 0,273587 |
| İstanbul Şehir Üniversitesi | 0,27432 |
| TED Üniversitesi | 0,274513 |
| Zirve Üniversitesi | 0,274868 |
| Mehmet Akif Ersoy Üniversitesi | 0,275118 |
| Beykent Üniversitesi | 0,27561 |
| Polis Akademisi | 0,275831 |
| Işık Üniversitesi | 0,276521 |
| İstanbul Sabahattin Zaim Üniversitesi | 0,278252 |
| Nişantaşı Üniversitesi | 0,278835 |
| Artvin Çoruh Üniversitesi | 0,279552 |
| İnönü Üniversitesi | 0,281342 |
| Yeditepe Üniversitesi | 0,281463 |
| Hasan Kalyoncu Üniversitesi | 0,281925 |
| Hakkari Üniversitesi | 0,282413 |
| Afyon Kocatepe Üniversitesi | 0,282534 |
| Ondokuz Mayıs Üniversitesi | 0,282761 |
| Uludağ Üniversitesi | 0,283182 |
| Düzce Üniversitesi | 0,285696 |
| Bülent Ecevit Üniversitesi | 0,286696 |
| Şifa Üniversitesi | 0,286878 |
| İstanbul Ticaret Üniversitesi | 0,287033 |
| Çankırı Karatekin Üniversitesi | 0,28719 |
| Galatasaray Üniversitesi | 0,290567 |
| Melikşah Üniversitesi | 0,2914 |
| Başkent Üniversitesi | 0,293059 |
| Yüzüncü Yıl Üniversitesi | 0,293074 |
| Abant İzzet Baysal Üniversitesi | 0,293372 |
| Kahramanmaraş Sütçü İmam Üniversitesi | 0,293694 |
| Hitit Üniversitesi | 0,295838 |
| Yaşar Üniversitesi | 0,296681 |
| Çankaya Üniversitesi | 0,298564 |
| Bezmiâlem Üniversitesi | 0,298759 |
| Osmaniye Korkut Ata Üniversitesi | 0,298968 |
| Kocaeli Üniversitesi | 0,298999 |
| Okan Üniversitesi | 0,299923 |
| Harran Üniversitesi | 0,300527 |
| Yıldırım Beyazıt Üniversitesi | 0,30192 |
| Adnan Menderes Üniversitesi | 0,302705 |
| Aksaray Üniversitesi | 0,303215 |
| Doğuş Üniversitesi | 0,303965 |
| İstanbul Teknik Üniversitesi | 0,304313 |
| Fatih Sultan Mehmet Üniversitesi | 0,304353 |
| İzmir Kâtip Çelebi Üniversitesi | 0,304853 |
| Mustafa Kemal Üniversitesi | 0,305983 |
| Celal Bayar Üniversitesi | 0,306376 |
| Piri Reis Üniversitesi | 0,307604 |
| Ordu Üniversitesi | 0,30815 |
| Bilecik Şeyh Edebali Üniversitesi | 0,309392 |
| Çukurova Üniversitesi | 0,309425 |
| Adana Bilim ve Teknoloji Üniversitesi | 0,310139 |
| Atatürk Üniversitesi | 0,312275 |
| Haliç Üniversitesi | 0,315196 |
| Eskişehir Osmangazi Üniversitesi | 0,318264 |
| İstanbul Aydın Üniversitesi | 0,320696 |
| Kırıkkale Üniversitesi | 0,322057 |
| Pamukkale Üniversitesi | 0,323109 |
| Mersin Üniversitesi | 0,326434 |
| İzmir Yüksek Teknoloji Enstitüsü | 0,327099 |
| İstanbul Kültür Üniversitesi | 0,332704 |
| Çanakkale Onsekiz Mart Üniversitesi | 0,333186 |
| Gaziosmanpaşa Üniversitesi | 0,334426 |
| Gaziantep Üniversitesi | 0,33451 |
| Özyeğin Üniversitesi | 0,336468 |
| Balıkesir Üniversitesi | 0,336587 |
| İstanbul Bilgi Üniversitesi | 0,34033 |
| Kadir Has Üniversitesi | 0,341905 |
| Akdeniz Üniversitesi | 0,342613 |
| Koç Üniversitesi | 0,346768 |
| TOBB Ekonomi ve Teknoloji Üniversitesi | 0,351168 |
| Marmara Üniversitesi | 0,354027 |
| Orta Doğu Teknik Üniversitesi | 0,358515 |
| Selçuk Üniversitesi | 0,367094 |
| Ege Üniversitesi | 0,375717 |
| Süleyman Demirel Üniversitesi | 0,380942 |
| Bahçeşehir Üniversitesi | 0,385826 |
| Dokuz Eylül Üniversitesi | 0,386229 |
| Yıldız Teknik Üniversitesi | 0,397495 |
| Ankara Üniversitesi | 0,398226 |
| Hacettepe Üniversitesi | 0,411034 |
| Boğaziçi Üniversitesi | 0,428301 |
| Sakarya Üniversitesi | 0,42942 |
| Bilkent Üniversitesi | 0,431946 |
| Gazi Üniversitesi | 0,443798 |
| İstanbul Üniversitesi | 0,444084 |
| Anadolu Üniversitesi | 0,449508 |

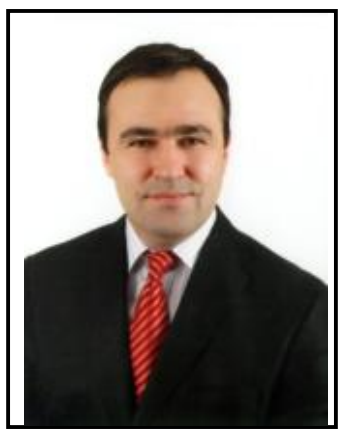

**Mehmet Lütfi Arslan** was born in Vezirkopru in 1972. He completed his undergraduate business degree in Marmara University of Istanbul. He obtained his MA and PhD in Social Sciences Institute of Marmara University. During his postgraduate studies, he worked in private sector. He is currently Assistant Professor of Management and Organization in newly founded Istanbul Medeniyet University and studies on management, organizations, entrepreneurship, and human resources management.

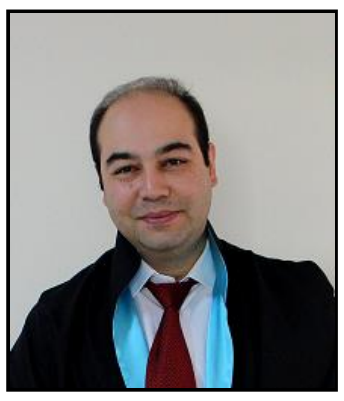

**Şadi Evren Şeker** was born in Istanbul in 1979. He has completed his BSc., MSc. and PhD. Degrees in computer science major. He also holds an M.A. degree in Science Technology and Society. His main research areas are Business Intelligence and Data Mining. During his post-doc study, he has joined data mining research projects in UTDallas. He is currently Asst. Prof. in Istanbul Medeniyet University, Department of Business. He is a senior member of IEDRC.